\documentclass[conference]{IEEEtran}
\IEEEoverridecommandlockouts

\usepackage{cite}
\usepackage{amsmath,amssymb,amsfonts}
\usepackage{algorithmic}
\usepackage{graphicx}
\usepackage{textcomp}
\usepackage{xcolor}
\def\BibTeX{{\rm B\kern-.05em{\sc i\kern-.025em b}\kern-.08em
    T\kern-.1667em\lower.7ex\hbox{E}\kern-.125emX}}
\begin{document}

\title{Agile TLB Prefetching and Prediction Replacement Policy}



\author{
\IEEEauthorblockN{Melkamu Mersha\IEEEauthorrefmark{1}, Tsion Abay\IEEEauthorrefmark{2}, Mingziem Bitewa\IEEEauthorrefmark{3}, Gedare Bloom\IEEEauthorrefmark{1}}
\IEEEauthorblockA{\IEEEauthorrefmark{1}\textit{ University of Colorado Colorado Springs}, Colorado Springs, USA \\}
\IEEEauthorblockA{\IEEEauthorrefmark{2}\textit{SOS HGS}, BahrDar, Ethiopia \\}
\IEEEauthorblockA{\IEEEauthorrefmark{3}\textit{University of Gondar}, Gondar, Ethiopia \\}
}


\maketitle

\begin{abstract}

Virtual-to-physical address translation is a critical performance bottleneck in paging-based virtual memory systems. The Translation Lookaside Buffer (TLB) accelerates address translation by caching frequently accessed mappings, but TLB misses lead to costly page walks. Hardware and software techniques address this challenge. Hardware approaches enhance TLB reach through system-level support, while software optimizations include TLB prefetching, replacement policies, superpages, and page size adjustments. Prefetching Page Table Entries (PTEs) for future accesses reduces bottlenecks but may incur overhead from incorrect predictions. Integrating an Agile TLB Prefetcher (ATP) with SBFP optimizes performance by leveraging page table locality and dynamically identifying essential free PTEs during page walks.
Predictive replacement policies further improve TLB performance. Traditional LRU replacement is limited to near-instant references, while advanced policies like SRRIP, GHRP, SHiP, SDBP, and CHiRP enhance performance by targeting specific inefficiencies. CHiRP, tailored for L2 TLBs, surpasses other policies by leveraging control flow history to detect dead blocks, utilizing L2 TLB entries for learning instead of sampling. These integrated techniques collectively address key challenges in virtual memory management.

\end{abstract}



\section{INTRODUCTION}
Virtual to-physical address space translation is the performance bottleneck in paging-based virtual memory systems. Translation Lookaside Buffer (TLB) is used for virtual-physical translation to overcome this problem. Modern computer architectures include hardware and software schemes to decrease address translation problems. A page table, software sided, stores the mappings of the virtual physical and physical address and is managed by the operating system. For example, the four-level radix tree page tables are employed in modern x86 architecture \cite{b1}. On the other hand, the TLB and the memory management unit accelerate the address translation pr. TLB cache stores the most recent translations for fast address retrieval. If some programs reference the virtual memory address, the searching process for this address starts inside the CPU to get fast retrieval. Therefore, the CPU is looking for translations inside the TLB; if the translation is missed in TLB, the CPU may traverse the entire page table one by one to obtain the missing translation. However, this condition incurs considerable latency and extra memory access costs because of sequential access of memory hierarchies \cite{b1}. Therefore, address translation significantly impacts memory access, particularly in low locality and workloads with large memory footprints. \\
\indent{}Researchers have recently proposed several techniques to mitigate this address translation challenge. We can classify these techniques into two categories as hardware and software schemes. The hardware scheme improves the TLB reach by establishing hardware and operating system support. There are various approaches in the software scheme categories, such as prefetching, page sizes and Superpages, and TLB replacement policies. Hardware and software schemes with multilevel Translation Lookaside Buffer hierarchies' implementations are the most common approaches for facilitating the address translation process. However, this approach also has a downside in performance [2].\\
\indent{}Numerous studies have been conducted on the significance of TLB performance and miss handlings for the overall system performance [2], [3]. The software or the hardware side handles the TLB misses. In the case of hardware handling of TLB miss, the CPU navigates the page table to obtain the requested PTE. If the requested PTE entry is found TLB hit occurs, and the CPU establishes the next address translation. If the requested PTE is not found, the CPU gets the page fault, and the operating system controls the problem. In the case of software handling of TLB miss, the page fault handler controls the issue. Handling the TLB miss on either side (hardware or software) compromises the execution time. The performance of the TLB can be improved at the application or operating system stage, including the compiler and optimization by the software side. Optimization can decrease the amount of PTE in the TLB by improving the locality. The structure of TLB size, associativity, and multilevel hierarchies also significantly affect the hardware side's miss rate and access time. \\
\indent{}This survey focuses on TLB Prefetching and TLB Predictive Replacement Policy. Prefetching the PTEs for future TLB access improves the performance of the TLB. TLB prefetching operates at the microarchitecture level and is not dependent on the system state. TLB prefetching also depends on memory access and does not disturb the current virtual memory system. The TLB prefetching system prompts the page walk to prefetch the PTEs. Each prefetching involves walking through the page table. However, prefetching also becomes an expensive approach if prefetching is not correct and not used by future TLB access. Hence, the prefetching may also trouble the address translation performance even if it reduces the TLB misses. Therefore, some unified solution is required to overcome this problem. Sample-Based Free TLB Prefetching (SBFP) can be integrated with any TLB prefetching technique to exploit the page table locality to improve TLB performance.
Furthermore, predictive cache replacement policies are used to improve TLB performance. LRU replacement policy predicts the near-instant reference interval on cache misses or hits [11]. Recent work shows that various researchers have employed several predictive replacement policies. Static re-reference interval prediction, global history reuse prediction, signature-based hit prediction, sampling-based dead block prediction, and control-flow history reuse prediction are the most common predictive policies. Specifical, this survey focuses on Agile TLB Prefetching (ATP) and Control-flow History Reuse Prediction (CHIRP) TLB improving performance methods.\\
\indent{} Finally, we recommended future work to improve the TLB performance. Multi-headed attention Recurrent Neural Network works better for the TLB prediction replacement policy.

\section{BACKGROUND}

\subsection{VIRTUAL MEMORY SUBSYSTEM}
Virtual to-physical address space translation is the performance bottleneck in paging-based virtual memory systems because every memory operation needs a translation process. TLB stores the frequently utilized address translations. The CPU searches the translation in the TLB for every access of memory, either instruction or data. The required translation is sent to the CPU if the TLB hit succeeds, but if the TLB miss occurs, the CPU walks through the page table one by one until the requested translation is found. This process results in latency. Furthermore, page fault happens if the request translation is not found in the page table. Hence, the amount of TLB misses and the TLB miss penalty directly affect CPU performance.\\
In the last recent years, various TLB performance improvement techniques have been developed by different researchers. However, TLB prefetching, replacement policies, page sizes, and Superpages are the most common approaches on the software side. Therefore, several approaches are incorporated and implemented to enhance the TLB performance Under each category.
\subsection{TLB Prefetching }
The CPU traverse of the page table is the cause of TLB misses. Therefore, the TLB misses and page table walks incur significant latency on the system performance. TLB prefetching of PTE for future access needs is a practical method to overcome this TLB miss latency. As recent research showed, several TLB prefetching techniques have been designed and implemented from the perspective of caches. In the TLB prefetching technique, there are various prefetching schemes; for example, a prefetch queue (small buffer) stores the prefetched PTE to keep the stability of the TLB content [4]. We explained some of the most common prefetching mechanisms in the following section.
\subsubsection{Sequential Prefetching (SP)}
Sequential prefetching techniques prefetches PTE sequentially. The next page table entry is accessed based on the existing reference \cite{b5}. 
 The most efficient way is prefetching the page table entry by triggering each requested fetch on every TLB hit to a prefetch queue. Another scheme, based on the TLB hit success rate, dynamically changes the size of prefetched page table entries. If the TLB miss occurs, the translation occurs on the prefetch buffer, and the prefetching process is triggered for the following address translation. Then the CPU continues immediately when the requested page table translation is delivered. On the other hand, if the prefetch buffer hit occurs, the page table entry is transferred into the TLB, the CPU quickly consumes the translation, and the prefetching is triggered for the following address translation process.

\subsubsection{Distance Prefetching (DP)}
Distance prefetcher is also a table-based TLB prefetcher mechanism. DP functions by keeping track of the spatial separation (distances between the consecutive address) mechanism [5]. DP generates a sequence of TLB misses by joining the TLB miss patterns and the distances of the virtual pages. The DP prediction table has three fields: the index tagged by the resultant distance, the current, and the previous page distance. If TLB miss occurs, DP calculates the current and the previous missing page distances, and the prediction table hits succeed, DP approves the prefetching process. Otherwise, the prediction table is updated by a new page table entry.

\subsubsection{Arbitrary Stride Prefetching (ASP)}
In real-world problems, many applications have regular-based stepwise patterns that sequential mechanisms cannot manage. For example, ASP TLB prefetcher uses a table to store different stride patterns, and the Program Counter indexes the table (PC) \cite{b2}, \cite{b5}. The table's index is used for reference prediction—each row of the table is labeled by four fields: index, stride, state, and the previous page. If the TLB miss happened, ASP checks the table for likely hits, and if the prediction table hits, the stride field gets updated based on the previous and the current missing pages. If the prediction table miss occurs, the state field gets reset, and the stride field is ignored. Finally, the state field counter is updated if the stride field does not change [2]. 

\subsubsection{Recency Based Prefetching (RP)}
The Recency-Based Prefetching approach functions on the principle that "pages referenced at around the same time in the past will also be referenced at around the same time in the future" [5]. RP attains this principle by building an LRU stack of PTEs. The ordering of the TLB access information is kept in the LRU stack. The top of the LRU stack stores the most recent referenced address translations, and the bottom stores the least recent referenced address translations. Hence, RP is the linked list structure system. If an entry is removed/evicted in the TLB, the entry is placed in the stack. The evicted entry's next pointer is pointed to the initial entry. Because of this, each node has two pointers in the page table. RP needs more space compared to SP and ASP, and it increases the size of the page table. If the PTE is loaded into the TLB miss, the RP system prefetches the PTEs considering the previous and the following pointers and stores them into the prefetch buffer for future demand.

\subsubsection{Markov Prefetching (MP)}
Markov Prefetching is a prediction table-based mechanism. The virtual page address is used to index the prediction table. Each entry of the prediction table has three fields: the indexing, the other two fields store the pages of the page table entries [2], [5].

\subsection{TLB Predictive Replacement Policy}
Predictive cache replacement policies are used to improve TLB performance. For example, the LRU replacement policy predicts the near-instant reference interval on cache misses or hits [11].
\subsubsection{Static Re-Reference Interval Prediction (SRRIP)}
SRRIP predicts the probability of the blocks that will be referenced again in the cache line [11]. Every block has 2-bit reference prediction values, and the re-reference prediction value places the block into one of four classes. The classes are arranged from the near-immediate to the distant re-reference range. The block is reused or replaced after the prediction is performed on the block placement or revision. Block with the distant re-referenced prediction is evicted, and the re-reference prediction value is incremented.

\subsubsection{Sampling-based Dead Block Prediction (SDBP)}
SDBP learns the access and the eviction pattern from the Sampler [1], [13]. The instruction is hashed to the prediction table when the load or store operation accesses a last-level cache. The counters are read out from the prediction tables and summed up, and the sum is thresholded to predict that the block is dead or live for the future. Signature-based Hit Prediction (SHiP) improves this idea of SDBP by reducing the number of predictions [10]..
\subsubsection{Global History Reuse Prediction (GHRP)}
GHRP is a predictive policy for branch history tables and i-cache placement [13]. GHRP and SHiP have similar structures, but in GHRP, the signature is used to index the instruction prediction table. GHRP uses conditional branch outcomes and low-order bits from addresses to produce an index. This index is used as a table of counters and stores records of reuse behavior.

\section{Related Work}
On the software side, the TLB performance can be improved with different mechanisms such as Prefetching, TLB replacement policy, page sizes, and Superpages. This survey provides high-level ideas about these TLB performance improvement techniques. Most recent works are summarized as follows.
\subsection{TLB Prefetching }
TLB misses highly affect the performance and energy of the system because of a page table walk to find the targeted address translation. Prefetching the page table entries for future TLB access reduces this performance bottleneck. However, every prefetching involves walking through the page table. Hence, TLB prefetching is also expensive and affects the system performance when the prefetching is incorrect, and the prefetched translation is not consumed by the processor [1]. Furthermore, recent research showed that no single TLB prefetches state-of-the-art functions best in all applications. In this section, we discussed some recent TLB performance-improving mechanisms to the best of our understanding.

\subsubsection{Sample-Based Free TLB Prefetching (SBFP)}
According to [1] recent work, the above problem can be solved by exploiting the page table locality. This exploiting locality can help to improve the TLB performance. The prefetching is accomplished by fetching adjacent cache-line page table entries for free. Therefore, this condition can be employed Sample-Based Free TLB Prefetching (SBFP) approach. SBFP is a dynamic technique that predicts the value of the free page table entries, and it prefetches only the most likely translations that can prevent TLB misses. At the end of every page walk, contiguous page table entries are stored in the same cache line to enhance TLB reach and minimize the page walk [6], [7]; however, this does not exploit the translation to decrease the prefetching cost [1]. The SBFP technique collects all adjacent page table entries into TLB prefetching queue when the page walk ends. The SBFP method can work with the combinations of any other TLB prefetchers and reduces the page table walk and utilization of address translation energy. SBFP considers the distance between the page table entries within the cache line, and PTE contains the required address translation and corresponding PTE to find for free distance. Therefore, many free possible distances exist depending on the targeted page table entry within the cache line.
\\

\includegraphics[scale=0.511]{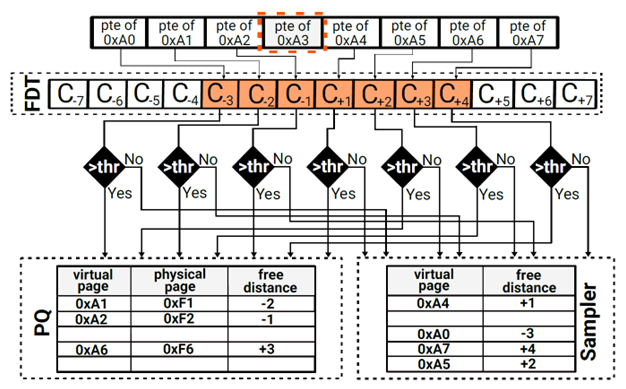}
\indent { \small Fig.1: Sampling-based free TLB prefetching mechanism [1].}
\\

SBFP scheme has three primary modules: Free Distance Table (FDT), Prefetching Queue, and Sampler, as figure 1 show the SBFP module functionality and components. The Sampler is a small buffer responsible for phase identification when free distances give valuable prefetches. The virtual page and its equivalent free distance are stored in each Sampler entry. The FDT decided whether the free page table should be placed into Sampler or Prefetching Queue. Every FDT counter controls the TLB hit ratio for each free distance, and Prefetching Queue stores the physical and virtual pages and the equivalent prefetches' free distance. The free distance regularly produces the PQ and Sampler hit, and the Sampler is explored only on Prefetching Queue misses. The SBFP modifies the FDT counter values based on Prefetching Queue or Sampler hits.

\subsubsection{Agile TLB Prefetcher (ATP)} 
SBFP can be integrated with any TLB prefetching technique to exploit the page table locality. SBFP can be integrated with any TLB prefetching technique to exploit the page table locality. Using the generic feature, we easily combine the SBFP and any TLB prefetcher; this is this work's novel idea. \\
ATP is employed as a decision tree structure. ATP enables the essential TLB prefetchers and disables non-essential TLB prefetchers dynamically by using adaptive selection and throttling approaches. ATP is composed of three prefetchers named P0, P1, and P2 and one PQ, as figure 2 shows. All prefetchers share the single PQ. The selection and throttling schemes also need another additional logic. This ATP additional logic has three units: one saturating counter ('enable\_pref') throttling scheme, two saturating counters ('select\_1 and select\_2'), and one Fake Prefetch Queue (FPQ) for each prefetcher (P0, P1, and P2). The 'select\_1' and 'select\_2' counters dynamically choose the correct TLB features. The FPQ contains the predicted virtual pages only. FPQ controls the accuracy of the prefetchers and updates' select\_1' 'select\_2', and 'enable\_pref' values. The SBFP and TLB prefetcher use the shared Prefetch Queue to store the prefetch requests.
\\

\includegraphics[scale=0.5]{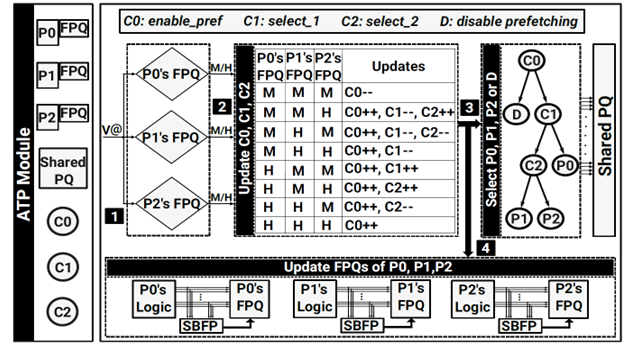}
\indent { \small Fig.2: ATP architecture adopted [1].}
\\
\\
\indent{}Agile TLB Prefetcher is designed with the benefits of SBFP. Specifically, ATP dynamically enables the applicable TLB prefetchers per TLB miss by relating patterns with various features [1]. ATP combines three low-cost TLB prefetchers efficiently and disables the TLB prefetching, which is not beneficiary in the execution phase.\\
a) Stride Prefetcher (STP): STP is the enhanced version of SP and uses the strides for the prefetching process [3].\\
b) H2 Prefetcher (H2P): H2P records the track of the previous two detected distances between the virtual pages, resulting in the TLB miss.\\
c) Modified Arbitrary Stride Prefetcher (MASP):  MASP is an advanced version of ASP [3]. ASP approves the prefetch request if two consecutive TLB hits exist in the prediction table. MASP modified this functionality in two different ways. MASP has three fields for each entry: a PC for index, stride, and the previous virtual page.\\

 \indent{} A combination of an Agile TLB Prefetcher and Sampling-Based Free TLB Prefetching mechanism is a unified solution to improve the address translation performance bottleneck. ATP exploits the locality of page table entries and decreases page walk memory references. SBFP dynamically prefetches the essential free PTEs per page walk and predicts feature TLB misses through sampling. This new combining approach reduces TLB misses by prefetching page table entries, which operate at the microarchitecture level. It is independent of a system state and does not disturb the existing virtual memory system. 
 
\subsection{TLB Reuse Prediction Policy } 
 \indent{} The larger the TLB size, the lower TLB misses but increasing the TLB size is problematic because it incurs access latency [8]. Furthermore, recent research shows that many programs spend more cycles for address translation due to missed TLB [9]. TLB replacement policy is one of the best solutions to improve the performance of TLB efficiency. This section focuses on a predictive replacement policy. Recent work shows that a reuse prediction policy can reduce TLB misses, and the TLB performance is improved with the prediction policy. The predictive replacement policy's primary goal is to predict the use of PTE before it is evicted. The predictive replacement policy is more innovative than the LRU replacement policy. Based on their functionality philosophy, the predictive policies are categorized into SRRIP, GHRP, SHiP, SDBP, and CHiRP. \\
 \indent{} Specifically, we presented the CHIRP, the most recent work [10]. Control-flow History Reuse Prediction used a history signature and replacement algorithm. The CHiRP TLB predictive replacement policy is good for L2 TLB. Because SRRIP, SHiP, and GHRP could not detect the dead blocks in the L2 TLB since they do not have the control flow history. CHIRP used L2 TLB entries for the learning process instead of the sampling method. CHIRP uses four basic features corresponding to reuse behavior: global path history of PCs, the current PC, conditional branch history, and unconditional address history, as the following figure shows.
\\
\includegraphics[scale=0.6]{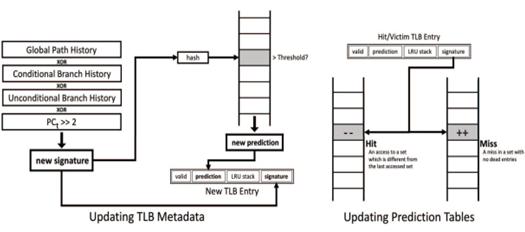}
\indent { \small Fig.3: CHiRP TLB metadata and prediction table [10].}\\
\\
 \indent{} The global path history has 64 bits width and is updated on each TLB access. The conditional and unconditional branch address histories are 64 bits wide. The conditional and unconditional branch address histories update the counter by shifting 8 bits. The current PC is always shifted to the right by using two bits. CHIRP uses these four features and correlates reuse history with the replacement policy using the CHiRP algorithm. The best correlations are combined into a signature. The signature uniquely indexes every TLB entry in the prediction table. The reuse of the TLB entry behavior is accessed by the signature based on the indexed prediction table. CHIRP keeps every L2 TLB entry metadata. The TLB Metadata contains three. The TLB metadata contains a prediction bit, three LRU position bits, and a signature bit. The prediction table is indexed with the hash function. The CHiRP uses this hashed prediction table for the prediction process. The entry is predicted as dead if the counter value is greater than the threshold. CHIRP updated the prediction table with negligible impact on TLB latency, whereas GHRP and SHiP updated the prediction table on every TLB access. CHIRP updated this prediction table with only a TLB miss based on the victim selection and prediction metadata update procedures. CHIRP operation on TLB miss includes victim selection, victim reuse history updating, and metadata prediction updating for every new TLB entry. CHIRP victim selection is made with entries predicted dead on the TLB miss, and CHIRP evicts the entry if no dead entry is found. The prediction table is updated when the victim is LRU. The effective eviction of LRU entry is good for the next dead candidate entry. The prediction table is indexed with the victim's entry signature to increase the counter value. The TLB metadata is updated after a new entry is inserted. The CHiRP uses this updated metadata for the following replacement decision process. \\
\indent{}On the TLB hit, the prediction table is indexed with the old signature, and the counter value of the table decrease by one. The old signature replaces the expected new signature. The prediction table is indexed again with this new signature, and the counter's value is reread. The decision of the entry, whether it is dead or alive for future use, is made based on the threshold value of the counter. The result of this prediction table status updates the metadata. CHIRP requires more access to the prediction metadata than other methods and updates the signature during both hit and insertion in the TLB.

\section{Future Research Directions}
SBFP is a dynamic technique that predicts the value of the free page table entries, and it prefetches only the most likely translations that can prevent TLB misses [1]. The SBFP technique collects all adjacent page table entries into TLB prefetching queue when the page walk ends. The Free Distance Table (FDT) decided whether the free page table entry should be placed into Sampler or Prefetching Queue. The Sampler is explored only on Prefetching Queue misses. Consider four fundamental concepts from the SBFP design principle: 1) the dynamic prediction, 2) adjacent PTEs collection into TLB prefetching queue, 3) the Sampler prefetching queue misses, and 4) the FDT decision process. SBFP uses some traditional algorithms for these fundamental concepts. However, deep neural networks and explainability techniques can easily enhance these four fundamental concepts (Recurrent Neural Networks, RNN)[15-16]. RNN is best known for sequential inputs since all adjacent page table entries are into TLB prefetching queue when the page walk ends. Second, The Sampler is explored only on Prefetching Queue misses means that RNN can quickly identify using either 0 or 1. Third, the FDT used the decision tree approach, but the decision tree is slow for extensive inputs. In this situation, RNN best fits a fixed and variable input length and vast input datasets. RNN can process all PTEs and quickly learn the behaviors of the entries rather than using sampling and explain it [17].\\
Combining an Agile TLB Prefetcher and a Sampling-Based Free TLB Prefetching mechanism is a unified solution to improve the TLB performance [1]. From this model architecture, we observed that ATP comprises three prefetchers named P0, P1, and P2 and one shared Prefetching Queue (PQ). Multi-headed attention RNN best fits this condition. The three independent prefetchers work independently (multi-headed), merge the results, and get the best results. The Fake Prefetch Queue (FPQ) contains the predicted virtual pages only. FPQ controls the accuracy of the prefetchers and updates. Hence, the three or more SBFPs work independently (multi-headed), and updating the (FPQ) is the second approach using multi-headed attention RNN. RNN performs better for FPQ controls the accuracy of the prefetchers and updates. Therefore Multi-headed attention RNN can improve the performance of the SBFP and an Agile TLB Prefetcher unified architecture. The RNN can also employ in the Control-Flow History Reuse Prediction policy approach.

\section{Conclusion}

This survey reviews TLB performance-improving techniques. TLB performance can be enhanced using the hardware side or software side. On the software side, we reviewed the TLB prefetching and prediction replacement policy techniques. Combining an Agile TLB Prefetcher and Sampling-Based Free TLB Prefetching approach is a unified solution to improve the TLB performance. SBFP dynamically prefetches essential free PTEs and predicts the future TLB misses, and ATP exploits the locality of page table entries. The reuse prediction replacement policy predicts that the cache item can be used again before eviction. Several policies exist, such as SRRIP, GHRP, SHiP, SDBP, and CHiRP. Specifically, we reviewed CHiRP, which is recent work. The CHiRP TLB predictive replacement policy is employed for L2 TLB because SRRIP, SHiP, and GHRP do not detect dead blocks in the L2 TLB since they do not keep to the control flow history.
Furthermore, CHIRP uses L2 TLB entries for the learning process instead of the sampling method since sampling does not work in the L2 TLB. Finally, we presented future work ideas to improve the TLB performance. We observed that the unified solution uses the TLB miss sampling technique, and CHiRP also uses the prediction table and control flow history for the learning process to make a prediction. The TLB performance may improve if we use LSTM neural networks in these situations. LSTM neural networks easily keep control flow history, prefetches the entries, and make an accurate prediction.

\end{document}